\documentclass{article}

\usepackage{arxiv}

\usepackage[utf8]{inputenc} 
\usepackage[T1]{fontenc}    
\usepackage{hyperref}       
\usepackage{url}            
\usepackage{booktabs}       
\usepackage{amsfonts}       
\usepackage{nicefrac}       
\usepackage{microtype}      
\usepackage{lipsum}		
\usepackage{graphicx}
\usepackage{natbib}
\usepackage{doi}

\title{Methods for Integrating Trials and Non-Experimental Data to Examine Treatment Effect Heterogeneity}

\author{
 Carly Lupton Brantner \\
  Department of Biostatistics\\
  Johns Hopkins Bloomberg School of Public Health\\
  \texttt{clupton1@jhu.edu} \\
   \And
   Ting-Hsuan Chang \\
   Department of Biostatistics \\
   Columbia Mailman School of Public Health \\
   \And
   Trang Quynh Nguyen \\
 Department of Mental Health\\
 Johns Hopkins Bloomberg School of Public Health
  \And
  Hwanhee Hong\\
  Department of Biostatistics and Bioinformatics\\
  Duke University \\
  \And
  Leon Di Stefano \\
  Department of Biostatistics\\
  Johns Hopkins Bloomberg School of Public Health\\
  \And
  Elizabeth A. Stuart \\
  Department of Biostatistics\\
  Johns Hopkins Bloomberg School of Public Health
 \\
}

\renewcommand{\shorttitle}{\textit{arXiv} Template}

\hypersetup{
pdftitle={Methods for Integrating Trials and Non-Experimental Data to Examine Treatment Effect Heterogeneity},
pdfsubject={stat-ME},
pdfauthor={Brantner et al.},
pdfkeywords={Treatment effect heterogeneity, Combining data, Generalizability and reproducibility},
}

\begin{document}
\maketitle

\begin{abstract}
Estimating treatment effects conditional on observed covariates can improve the ability to tailor treatments to particular individuals. Doing so effectively requires dealing with potential confounding, and also enough data to adequately estimate effect moderation. A recent influx of work has looked into estimating treatment effect heterogeneity using data from multiple randomized controlled trials and/or observational datasets. With many new methods available for assessing treatment effect heterogeneity using multiple studies, it is important to understand which methods are best used in which setting, how the methods compare to one another, and what needs to be done to continue progress in this field. This paper reviews these methods broken down by data setting: aggregate-level data, federated learning, and individual participant-level data. We define the conditional average treatment effect and discuss differences between parametric and nonparametric estimators, and we list key assumptions, both those that are required within a single study and those that are necessary for data combination. After describing existing approaches, we compare and contrast them and reveal open areas for future research. This review demonstrates that there are many possible approaches for estimating treatment effect heterogeneity through the combination of datasets, but that there is substantial work to be done to compare these methods through case studies and simulations, extend them to different settings, and refine them to account for various challenges present in real data.
\end{abstract}


\maketitle

\section{Introduction}

Identifying the right treatment for the right patient can improve quality of healthcare for individuals and populations. Treatments for disorders and diseases like depression \citep{trivedi_evaluation_2006}, schizophrenia \citep{samara_how_2019}, and diabetes \citep{xie_precision_2018} can exhibit differential treatment effects across individuals due to \emph{effect moderators}, defined as known and unknown individual, genetic, environmental, and other characteristics that are associated with the effectiveness of medical treatments \citep{baron_moderatormediator_1986}. Finding ways to identify and leverage effect moderators at the point of care to facilitate clinical decision-making can improve efficiency, quality and outcomes of healthcare. 

Although crucial for delivery of treatment and preventative medicine, detecting treatment effect heterogeneity is challenging with common study designs. Randomized trials yield comparable treatment groups on average but are typically under-powered to detect moderation. One rule-of-thumb is that study samples need to be four times larger to test an effect moderator than to detect the overall average effect \citep{enderlein_fleiss_1988}. In addition, randomized trial samples are also often not representative of the target population for which treatment decisions will be made; for instance, Black individuals are on the whole underrepresented in pivotal clinical trials \citep{green_despite_2022}. Therefore, conclusions from one particular trial might not reflect conclusions for a target population, and different trials might give conflicting results due to differences in their enrolled participants. On the other hand, large-scale non-experimental studies can have improved external validity, but these studies can suffer from confounding bias. Given power concerns in single randomized trials and bias concerns in non-randomized studies, much can be gained by combining multiple trials, or combining experimental and non-experimental studies, to examine effect moderation \citep{berlin_individual_2002,brown_methods_2013}. 

Many methods have been proposed to examine effect moderation in a single study. One of the popular approaches is to prespecify a few key subgroups and fit models with treatment-subgroup interactions. This approach is limited in that data analysts could explore a range of possible subgroups and report only those that are statistically significant \citep{kent_assessing_2010}; additionally, this approach does not allow the contribution of multivariate factors in effect moderation. Another approach is ``risk modeling'' \citep{kent_assessing_2010,kent2020predictive}, where a risk score is created using the covariates to predict the outcome (usually outcome under the comparison/control condition), and the treatment effect is assessed based on the interaction between treatment and this risk score in a regression model of the outcome. This review focuses on what is sometimes called ``effect modeling''. Effect modeling spans a spectrum that includes parametric approaches in which a few effect moderators are pre-specified, and nonparametric approaches where effect moderation is assumed to be via some potentially complex function of a large set of covariates. Regression analyses and variable selection are common approaches for the former; machine learning methods for the latter.

In order to examine treatment effect heterogeneity based on observed characteristics, the target estimand in the present work is the conditional average treatment effect (CATE). Notation for this estimand is presented in the following section. The CATE is a general function of covariates that could be quite complex and so requires large sample sizes to estimate reliably. A key assumption when combining studies to estimate the conditional average treatment effect is that the CATE function is substantially similar across studies. When discussing the CATE, it is relevant to note that the CATE function is related to subgroup average treatment effects and identification of groups who benefit from treatment; these similar goals are mostly outside of the scope of this review. We therefore focus on the CATE and mention subgroup treatment effects and other similar topics briefly when relevant.

There have been recent statistical advances in modeling heterogeneous treatment effects and a separate burgeoning interest in combining data from multiple sources. A select few works have done both – simultaneously leveraging data from multiple studies to assess treatment effect heterogeneity. Methods like these are needed to best harness the available data to optimize and individualize treatments, and to leverage information from multiple studies to provide more systematic, comprehensive, and generalizable conclusions. This paper reviews these novel methods of assessing treatment effect heterogeneity using multiple studies in the form of multiple randomized trials, or one randomized trial with a large observational dataset. We focus on methods identifying which of two treatments is more likely to improve outcomes for an individual or subgroup – a causal question that sits at the core of clinical practice. In this review, we consider the situation where the variables are similarly defined and available from all studies. It is common though that different studies may have different sets of variables. In this more complicated case, either harmonization is needed on the variables or some shared structure is required on conceptually related variables. We will return to this point in the Discussion section (\ref{futuredir}).

Methods discussed in this paper are broken down based on data setting: aggregate-level data, federated learning, and individual participant-level data (IPD). The aggregate-level data setting occurs when researchers only have access to summary information from each study. With aggregate-level data, individual-level effect heterogeneity can only be truly assessed if each study estimated treatment-covariate interactions using the same statistical models (e.g., same link function, same set of covariates), which is not often feasible. In the federated learning setting, sensitive individual-level data are distributed across decentralized studies and cannot be shared beyond their original storage location \citep{vo_federated_2021}. Finally, the IPD setting is the most straightforward and powerful scenario for assessing treatment effect heterogeneity, as individual-level covariates are available from all studies simultaneously. With IPD, we can harmonize covariates, estimate effect moderation by using the same statistical models in each study, and assess model assumptions consistently. 

Within each of these data settings, methods are primarily geared towards either combining multiple RCTs or one RCT with one observational dataset. We discuss the use of meta-analysis models with multiple RCTs \citep{debray_get_2015, burke_meta-analysis_2017}, along with the opportunity to employ variable selection approaches to identify effect moderators \citep{seo_comparing_2021}. When combining an RCT with observational data, we consider various methods that allow for complicated relationships to be included in the treatment effect function and account for potential bias from the observational data. These methods can involve estimating the CATE in the RCT and observational data separately and then combining them through an estimated weighting factor \citep{rosenman2022propensity, rosenman_combining_2020, cheng_adaptive_2021,yang_elastic_2020}, or estimating the observational CATE and the confounding effect in the observational dataset \citep{kallus_removing_2018, yang_elastic_2020,wu2021integrative,hatt2022combining}. \cite{colnet_causal_2021} reviewed some methods that combine RCT and observational data, and we extend upon this review by focusing on this combination explicitly for treatment effect heterogeneity. We also add in more methods that combine RCT with observational data along with methods that focus on combining multiple RCTs. In general, there are many approaches outside of those we reference here that focus on estimating the \textit{average} treatment effect by combining datasets, some of which are discussed by \citeauthor{colnet_causal_2021} (\citeyear{colnet_causal_2021}); we choose to primarily focus on efforts to examine treatment effect heterogeneity in the present review.

To provide context to the methods discussed in this review, we can consider a few example scenarios. We first consider an assessment of the efficacy of surgery in stage IV breast cancer according to 15 studies where researchers combining the studies only had access to aggregate-level data \citep{petrelli2012surgery}. We also discuss a comparison of outcomes for veterans who received the Moderna versus the Pfizer vaccination for COVID-19 in five different sites where IPD was available within each site but could not be shared across sites, known as a ``federated learning'' situation \citep{han_federated_2021}. Another setting investigates a diabetes medication, pioglitazone, versus placebo for individuals coming from one of six RCTs, where IPD was available in each trial \citep{hong2015incorporation}. And finally, we discuss data assessing the treatment effect comparing two active treatments for major depression, duloxetine and vortioxetine, wherein we have access to IPD from a combination of RCT data and electronic health records (EHR) from a hospital system \citep{combining_present}. These scenarios all could clearly benefit from combining data to examine heterogeneity in treatment effects, but they each require distinct considerations and statistical approaches to best integrate information. We will use these examples throughout the paper to ground the methods in specific applications.

Importantly, to effectively combine information from multiple datasets, the original studies need to have high transparency and reproducibility. Whether data are reported in aggregate or at the individual participant level, researchers using the data for additional analyses -- such as those discussed here -- need extensive information about how the data were collected, analyzed, and presented to be able to determine if and how to combine the information with other datasets. It is therefore vital to keep these ideas of transparency and reproducibility of data, code, and results at the forefront when applying these methods. Movements towards data sharing and reproducible research will greatly facilitate the types of research discussed here, which can lead to important new insights regarding effect heterogeneity that cannot be answered from single studies alone due to generalizability, sample size, or confounding concerns.

In the following section (\ref{notation}), we introduce the estimand and assumptions. The next sections are then organized based on the level of data access so that researchers can determine available methods in their given data setting. Specifically, Section \ref{AD} discusses aggregate-level data; Section \ref{FL}, federated learning; and Section \ref{IPD}, individual participant-level data (IPD). Finally, Section \ref{futuredir} compares methods and provides an overview of potential future areas for research.

\section{Notation} \label{notation}

\subsection{Target Estimand}
Our target estimand to assess effect heterogeneity is the conditional average treatment effect (CATE), defined using the potential outcomes framework under the Stable Unit Treatment Value assumption \citep{rubin_estimating_1974}. Suppose $S$ is the categorical variable indicating study membership, $A = 0, 1$ is a binary treatment variable, $Y$ is the observed outcome, $Y(1)$ and $Y(0)$ are the potential outcomes under treatment and control respectively, $\boldsymbol{X}$ is a set of covariates, and $\boldsymbol{Z}$ is a subset of $\boldsymbol{X}$ containing the proposed effect moderators.

The CATE can be formally defined as a function of $\boldsymbol{X}$:
$$\tau(\boldsymbol{X})=\ g(E[Y(1)|\boldsymbol{X}]) - g(E[Y(0)|\boldsymbol{X}])$$ \citep{abrevaya_estimating_2015,kunzel2019metalearners}, where $E[.|.]$ denotes conditional expectation in the target population of interest and $g(.)$ is a link function that defines the scale on which the interactions occur, whether additive (mean or risk difference) or multiplicative (risk, rate, or odds ratio). In this paper we primarily discuss a continuous outcome, in which case we use the identity link function and write the CATE as 
\begin{equation}  \label{CATE}
    \tau(\boldsymbol{X})=\ E[Y(1)- Y(0)|\boldsymbol{X}].
\end{equation}
This $\tau(.)$ can often be assumed to be a flexible function in which all covariates are considered as potential moderators, so we do not have to a priori differentiate $\boldsymbol{Z}$ and $\boldsymbol{X}$ when methods allow for this flexibility.

One can also consider study-specific CATE functions. This is often the case when researchers are interested in assessing heterogeneity of the treatment effect functions across trials/datasets, or when this heterogeneity is high and it is potentially unreasonable to combine information across studies. We can denote study by $S$: in the case where data is being combined from one RCT and one observational dataset, $S=0$ will indicate RCT and $S=1$ observational data; otherwise, $S$ will be a categorical variable ranging from $1$ to $K$, where $K$ is the number of RCTs. The above Equation (\ref{CATE}) defines a general CATE that is not study-specific. When estimating study-specific CATEs, Equation (\ref{CATE}) can be rewritten as 
\begin{equation} \label{CATE_nonpara_study}
    \tau_s(\boldsymbol{X}) = E[Y(1)-Y(0)|\boldsymbol{X},S=s].
\end{equation}

In most of the methods to follow, the CATE is defined by conditioning on a set of available covariates, $\boldsymbol{X}$.  An alternative is to a priori define subgroups of interest and estimate subgroup-specific treatment effects. This approach is similar to the methods discussed in this review but somewhat distinct because subgroups must be specified first. The form of the estimand when examining subgroup-specific effect estimates is instead $$\tau_k = E[Y(1)-Y(0)|K=k]$$ where $K$ represents subgroup membership \citep{rosenman_combining_2020,rosenman2022propensity}.

\subsection{Assumptions}

Across many methods, the key assumption that allows pooling data from multiple studies to estimate the treatment effect is that either entire or partial components of the treatment effect function $\tau(\boldsymbol{X})$ is shared across studies. This review also focuses solely on the case when there are only two treatments (or one treatment and one control/placebo) being compared. If there are more than two conditions being compared, different approaches would need to be used (i.e., network meta-analysis;  \cite{efthimiou_getreal_2016,debray_overview_2018, hong2015incorporation}). Aside from these overarching assumptions, individual methods employ their own specific assumptions. When multiple RCTs are included in meta-analyses, they are often assumed to have similar eligibility criteria (specifically in terms of the covariates thought to be effect modifiers) \citep{dahabreh_towards_2020}, and distributional assumptions are made for model parameters \citep{debray_get_2015}.

Broadly, parametric approaches require the assumption of a parametric relationship between covariates (including treatment, effect moderators, and interactions between the two) and outcomes; further, this parametric relationship is assumed to be approximately correctly specified \citep{debray_get_2015,yang_improved_2022,yang_elastic_2020}. Specifically in the meta-analytic framework when combining multiple RCTs, effect moderation is often assessed using treatment-covariate interaction terms. This approach typically uses an outcome model of the form
$$h(E(Y)) = \mu(\boldsymbol{X}) + A\times\tau(\boldsymbol{Z}),$$ 
where $h(.)$ is a link function, $\mu(\boldsymbol{X})$ is the modelled mean of the outcomes under control, $\boldsymbol{Z}$ contains a subset of the variables in $\boldsymbol{X}$ that often needs to be pre-specified, and $\tau(\boldsymbol{Z})$ is the the CATE function: \begin{equation} \label{CATE_para_uni}
    \tau(\boldsymbol{Z}) = \delta+ \boldsymbol{\theta}^T\boldsymbol{Z}.
\end{equation} 
In this expression for $\tau(\boldsymbol{Z})$, $\delta$ corresponds to the effect of treatment $A$ when $\boldsymbol{Z}=0$ (or when the covariates in $\boldsymbol{Z}$ equal their means if they have been centered), and $\boldsymbol{\theta}$ corresponds to the coefficients of treatment-moderator interaction terms $A\boldsymbol{Z}$ in the $h(E(Y))$ model. Similarly to the general format of the CATE in Equation (\ref{CATE}), this parametric form of $\tau(\boldsymbol{Z})$ can be expressed as multiple study-specific functions:
\begin{equation}\label{CATE_para_study}
    \tau_s(\boldsymbol{Z}) = \delta_s + \boldsymbol{\theta}_s^T\boldsymbol{Z}.
\end{equation}

When combining an RCT with an observational dataset, there are a few within-study assumptions, including unconfoundedness (Assumption \ref{unconf}), positivity (Assumption \ref{positivity}), and consistency (Assumption \ref{consistency}) \citep{colnet_causal_2021, cheng_adaptive_2021}:

\newcommand{\indep}{\perp \!\!\! \perp}
\newtheorem{assumption}{Assumption}

\begin{assumption} \label{unconf}
$\{Y(0), Y(1)\} \indep A | \boldsymbol{X}$ within each study.
\end{assumption}
\begin{assumption} \label{positivity}
For almost all $\boldsymbol{X}$ with $\pi(\boldsymbol{X}) = P(A=1|\boldsymbol{X})$ (the propensity score), there exists a constant $c>0$ such that $c<\pi(\boldsymbol{X})<1-c$ within each study.
\end{assumption}
\begin{assumption} \label{consistency}
$Y = AY(1) + (1-A)Y(0)$ almost surely.
\end{assumption}

The unconfoundedness assumption (\ref{unconf}) is satisfied by design in an RCT. Assumption \ref{positivity} also holds by design in an RCT since the probability of treatment is independent of observed covariates and is pre-specified.

When combining datasets, we expand upon the previous assumptions. In the setting where observational data is being combined with an RCT, the unconfoundedness assumption (\ref{unconf}) can be relaxed in the observational data. This is because there are analysis possibilities with multiple datasets that include assessing whether this assumption is met or not and using the RCT to account for any confounding in the observational data \citep{cheng_adaptive_2021, yang_elastic_2020, yang_improved_2022}. Assumption \ref{consistency} in the multi-study setting implies that the treatments being compared are the same across all studies (since there is no $s$ subscript) to ensure that the potential outcomes $Y(0)$ and $Y(1)$ are well-defined. We also can introduce two other assumptions that are involved at some level in methods that combine an RCT with observational data; these assumptions include study membership positivity (Assumption \ref{studypositivity}) \citep{colnet_causal_2021,cheng_adaptive_2021} and unconfounded study membership (Assumption \ref{studyunconf}) \citep{hatt2022combining,cheng_adaptive_2021,kallus_removing_2018}.

\begin{assumption}\label{studypositivity}
 For almost all $\boldsymbol{X}$ there exists a constant $d>0$ such that $d<P(S=s|\boldsymbol{X}=\boldsymbol{x})<1-d$. 
\end{assumption}
\begin{assumption} \label{studyunconf}
 $\{Y(0), Y(1)\} \indep S | \boldsymbol{X}.$
\end{assumption}

The following sections break down methods based on available data.

\section{Aggregate-Level Data} \label{AD}

The broadest level of data access is in the form of aggregate-level data (AD), where individual studies have been carried out and analyzed, and only summary data (e.g., sample mean, standard deviation, or regression model coefficient estimates) are available. AD are often used in meta-analyses when IPD are unavailable. Meta-analysis with AD can estimate average effects effectively and provide similar results as meta-analysis with IPD \citep{burke_meta-analysis_2017,hong2015incorporation}. However, aggregation bias (also known as the ecological fallacy), which occurs when conclusions are incorrectly drawn about individuals when the relationship is found at the group level, can easily be introduced if researchers want to make a conclusion about individual-level effect moderation when only AD is available \citep{berlin_individual_2002,debray_get_2015,teramukai2004individual}. This aggregation bias will not be present if each paper reports subgroup-specific outcomes for all necessary subgroups; however, this is rare in practice because subgroups are often defined by more than one covariate. AD therefore has limited power for detecting effect moderation \citep{lambert_comparison_2002}. However, IPD is not always easy to access or use, so the following section discusses what can be done with AD. In framing this discussion, one can think of the example assessing the effects of tumor-removal surgery in individuals with breast cancer \citep{petrelli2012surgery} using aggregate data from several relevant studies.

\subsection{Meta-Analysis of Interaction Terms} \label{ma_of_inter}

If AD is all that is available for a question of interest, there is still an opportunity to estimate individual-level effect moderation under specific circumstances. If all previous studies have performed similar analyses and have included a particular treatment-covariate interaction term using the IPD from that given study, then these interaction terms can be pooled at the aggregate level \citep{simmonds_covariate_2007,kovalchik2013aggregate}. For instance, although this approach was not taken by \citeauthor{petrelli2012surgery} (
\citeyear{petrelli2012surgery}), if a treatment-age interaction term was estimated in each of the individual studies assessing the effect of surgery on mortality in individuals with stage IV breast cancer, then these interaction terms could be pooled together. In this way, researchers can estimate an individual-level effect moderation term across multiple studies and can combine such terms to estimate $\tau(\boldsymbol{Z})$ as in Equation (\ref{CATE_para_uni}). However, this requires that the studies assess and report the interactions of interest consistently. Similarly, the aggregate data could include subgroup-specific treatment effects rather than interactions, which could also be pooled to describe effect moderation if the effects are reported in each study \citep{godolphin2022estimating}.

\subsection{Meta-Regression}

If such study-specific interaction coefficients are not available across all studies, AD can be also modeled through meta-regression with treatment-covariate interaction terms, where  importantly only aggregate level covariates (e.g., mean age, proportion female) are available. For example, the individual-level covariate of interest might be whether the person has severe disease or not; in an AD meta-regression, this covariate would become the percentage of individuals in the study who have severe disease. Meta-regression was the approach taken by \citeauthor{petrelli2012surgery} (\citeyear{petrelli2012surgery}) in their assessment of surgery efficacy. Specifically, they investigated hazard ratios of overall survival according to the fifteen different studies and did so while including covariates such as median age and mastectomy rate.

AD analyses can handle study-level effect moderators well. However, the ability to assess individual-level moderators depends on the level of detail available in the AD. Multiple papers have assessed the differences between AD and IPD meta-regressions for estimating treatment effect heterogeneity. In an analysis by Berlin and colleagues (\citeyear{berlin_individual_2002}), models using IPD picked up on a key effect moderator that had been found in previous literature, but all models using AD missed this effect moderator at the group level. Extensive simulation studies also have shown that the power for detecting treatment effect moderation is much lower in meta-regression using AD; in these simulations, effect moderation was only effectively discovered in AD analyses when there were a large number of trials with large sample sizes \citep{lambert_comparison_2002}. Again, relationships that are picked up in an AD meta-regression cannot be immediately interpreted as individual-level effects; for example, if the percentage of individuals with severe disease is an effect moderator in the AD model, researchers cannot immediately conclude that the individual-level presence of severe disease is an effect moderator at the individual level. 

Furthermore, the aggregate-level covariates also often do not vary much across studies. Since studies included in meta-regressions require similar eligibility criteria, they likely will have somewhat similar covariate distributions. For instance, the percentage of individuals with severe disease is likely to be similar across trials; in this case, the interpretation of effect moderation cannot be extrapolated beyond the aggregate-level range of the covariates.

The estimand in meta-regression can still be considered to be a version of the CATE, but it is the CATE according to group-level effect moderators; for example, it could be written like Equation (\ref{CATE_para_uni}) but as $\tau(\bar{\boldsymbol{Z}})$ where $\bar{\boldsymbol{Z}}$ consists of aggregations of $\boldsymbol{Z}$ at the study-level. Such an estimand assumes that the included studies are representative of the target population of studies.

\section{Federated Learning} \label{FL}

Federated learning (similar to distributed modeling) uses a combination of IPD and AD; namely, IPD exists across decentralized studies but can only be accessed in the study in which it is stored \citep{yang2019federated}. An example of this is a study of the efficacy of two COVID-19 vaccinations (developed by Moderna and Pfizer) for preventing COVID-19 in veterans in five Veterans Affairs sites \citep{han_federated_2021}. This data setup is increasingly common in fields where there is interest in combining multiple cohorts (``cohort consortia''), but where data privacy concerns prohibit full direct data sharing. Therefore, the IPD data must be turned into AD or aggregated models so that information can be shared across studies.

We discuss two approaches for CATE estimation in federated learning in this section. Other approaches exist that focus on estimating the average treatment effect (ATE) \citep{han_federated_2021}, and those can be extended to CATE estimation but must provide sufficient information about the parameters of effect moderation. Depending on the ATE approach, it is unclear how easily the method can be extended to CATE estimation; we focus instead on methods explicitly focused on CATE estimation.

\subsection{Meta-Analysis after Local Model Formulation}

There are three steps in meta-analysis within the federated learning setting: (1) fit models within studies, (2) aggregate the model coefficients, and then (3) conduct a meta-analysis \citep{silva2019federated}. This is similar to the meta-analyses of interaction terms using aggregate data discussed in Section \ref{ma_of_inter}. A key difference here is that federated learning models apply a pre-determined statistical model including desired interaction terms so that the interaction effects are assessed consistently across all studies, while the traditional meta-analysis with AD has access to model coefficient estimates but not the model fitting process. Here, the estimand of interest is the common CATE function as in Equation (\ref{CATE_para_uni}) that is calculated by summarizing model coefficients corresponding to interaction terms $A\boldsymbol{Z}$ (treatment-moderator) and $A$ (treatment) from each study-specific regression.

\subsection{Tree-Based Ensemble}

Another option within federated learning would be to still create study-specific models first, but to use information from other studies to improve those individual models. \citeauthor{tan_tree-based_2021} (\citeyear{tan_tree-based_2021}) use tree-based ensemble methods to combine information about treatment effect heterogeneity from multiple separate studies. Specifically, they allow for study-level heterogeneity as well as heterogeneity due to individual-level covariates. 

Their procedure involves first fitting models to estimate the CATE in each of $K$ individual studies, using single-study machine learning methods like causal forests (see Supplementary Material \citep{brantnersupp}). These $K$ study-specific models are then applied to a single ``coordinating study'', so that each individual in the coordinating study has $K$ estimates of the CATE. In other words, if there are $n$ individuals in the coordinating study, there will be $n*K$ CATE estimates. Finally, these $n*K$ estimates are used as outcomes in an ensemble regression tree or random forest, in which the predictors are the individual-level covariates and an indicator of the study model from which the specific CATE estimate was estimated. Ultimately, this method provides study-specific CATE functions (Equation (\ref{CATE_nonpara_study})) that have hopefully been made more accurate because they have been adjusted to incorporate information from other studies. \citeauthor{tan_tree-based_2021} (\citeyear{tan_tree-based_2021}) applied this approach to investigate the effects of oxygen saturation on hospital mortality across 20 hospitals and found effects that varied across sites but did not have high levels of within-site heterogeneity based on covariates like age or gender.

\section{Individual Participant-Level Data} \label{IPD}

Finally, when individual participant-level data (IPD) is available from all studies, treatment effect heterogeneity can be estimated through a wide variety of methods. Recently, many novel methods have been proposed and are actively being developed. While the previous two settings of AD and federated learning are more restrictive, estimating individual-level effect moderation in this setting with all IPD available is much more feasible and flexible. The methods to follow are broken down based on whether the data being combined is from multiple RCTs or from one RCT and one observational dataset. Many of the methods in this multi-study setting build upon single-study methods, which are discussed in depth in the Supplementary Materials \citep{brantnersupp}.

\subsection{Combining Multiple RCTs}

As mentioned when discussing aggregate data, meta-analyses are an effective and widely used parametric approach for combining information from multiple RCTs \citep{riley2021individual}. Recently, more and more IPD has become accessible to researchers, allowing them to go a step further from AD and more effectively assess effect moderation. Having IPD available, such as in the example of assessing the effects of pioglitazone for individuals with diabetes \citep{hong2015incorporation}, allows for baseline individual-level covariates to be used to study subgroup effects and effect moderation at the individual level. 

\subsubsection{Types of IPD Meta-Analyses} \label{types}

There are two commonly discussed IPD meta-analysis estimation methods: two-stage and one-stage. In two-stage IPD meta-analysis, aggregate statistics are calculated within each study (e.g., overall treatment effects, effects for each subgroup, interaction terms), and then these results are combined in a between-study model. In one-stage IPD meta-analysis, all individual-level data are put directly into a hierarchical or multilevel model \citep{burke_meta-analysis_2017}. Although results with respect to average treatment effects are often similar between the two approaches \citep{burke_meta-analysis_2017,debray_get_2015,tierney_individual_2015}, model assumptions do differ, and choosing the approach that seems best fit to a specific research question is an important decision. In this paper, we focus on one-stage IPD meta-analysis because of its flexibility \citep{debray_get_2015}.

\subsubsection{One-Stage IPD Meta-Analysis} \label{ipdma}

In one-stage IPD meta-analysis, a common technique is to use a generalized linear mixed model (GLMM) to estimate the mean outcome given covariates. The model can have the form
\begin{equation}\label{glmm}
    g(E(Y_{is})) = \alpha_s + \delta_sA_{is} + \boldsymbol{\beta}_{s}^T\boldsymbol{X}_{is} + \boldsymbol{\theta}_{s}^TA_{is}\boldsymbol{Z}_{is},
\end{equation} 
where $Y_{is}$ is the outcome for individual $i$ from study $s$, $\alpha_s \sim N(\alpha, \sigma_{\alpha}^2)$ is a study-specific intercept, $\delta_s \sim N(\delta, \sigma_\delta^2)$ is the vector of study-specific treatment effects when the covariates are set to 0 (or their means, if centered), $\boldsymbol{\beta}_s \sim N(\boldsymbol{\beta}, \boldsymbol{\Sigma}_{\boldsymbol{\beta}})$ is the study-specific vector of main effects of covariates on the outcome, and $\boldsymbol{\theta}_s \sim N(\boldsymbol{\theta}, \boldsymbol{\Sigma}_{\boldsymbol{\theta}})$ is the study-specific vector of effect moderation terms \citep{seo_comparing_2021}. Here, $\sigma_\alpha^2$, $\sigma_\delta^2$ and the diagonal elements of $\boldsymbol{\Sigma}_{\boldsymbol{\beta}}$ and $\boldsymbol{\Sigma}_{\boldsymbol{\theta}}$ measure the between-study variability of the effects. $\beta_s$ and $\theta_s$ are often assumed to be uncorrelated in the literature; however, we can extend this model to allow for correlation between $\beta_s$ and $\theta_s$. 

If the outcome is continuous (as assumed in this paper), $g(.)$ is often set to be the identity function; if the outcome is binary, $g(.)$ could be the logit link function. Key parameters of interest are $\delta$, which indicates an overall measure of the treatment effect when the moderators are set to 0, and $\boldsymbol{\theta}$, which indicates the magnitude of the effect moderation. For easy interpretation, covariates can be centered at zero so that the treatment effects, $\delta_s$ represent the treatment effects at the mean value of each covariate \citep{dagne_testing_2016, gelman2020regression}.

The model above includes random effects for all coefficients, and so explicitly models between-study heterogeneity for each coefficient (the $\boldsymbol \beta_s $'s and $\boldsymbol \theta_s $'s). This approach can be thought of as interpolating between two extremes. The first of these is a ``no-pooling'' model, with the same structure as Equation (\ref{glmm}) but with study-specific coefficients fit as fixed effects independently to the data from each study. Such a model avoids the sharing of information across studies, but also includes more free parameters, which may be less stably estimated. This approach also does not ultimately provide a global treatment effect estimate across studies, as all studies are given their own fixed coefficients.

A simpler model would treat some coefficients as shared across studies. This might take the form of assuming a common intercept or slope \citep{thomas_systematic_2014}; for example, in Equation (\ref{glmm}), if between-study variability of the main covariate effects (represented by $\boldsymbol{\Sigma}_{\boldsymbol{\beta}}$) were small, a common coefficient could be estimated instead by replacing $\boldsymbol{\beta}_s$ with $\boldsymbol{\beta}$. In practice, $\boldsymbol{\theta}$ is often assumed to be shared across studies. GLMMs can quickly become too complicated if many effects are allowed to vary across studies (especially when study sample sizes are small); on the other hand, the model might be misspecified if it ignores important variation that does exist. Therefore, each coefficient -- and whether it should be treated as common across studies, modelled as random, or estimated independently within each study -- should be considered carefully to ensure that the model effectively represents between-study variability while still being sufficiently simple. 

GLMMs can be fit under both frequentist and Bayesian frameworks \citep{debray_get_2015}. If a Bayesian framework is used, prior distributions need to be assigned to each parameter; an option for this is non-informative priors to all parameters of interest \citep{mccandless2009bayesian}. Informative priors can be used when information about the parameters is available from expert opinion or historical data analysis. \citeauthor{hong2015incorporation} (\citeyear{hong2015incorporation}) utilize a Bayesian framework for their analysis of diabetes medication; however, they compare more than just two treatments and perform network meta-analysis, which is not the focus of this paper.

One other consideration in one-stage IPD meta-analysis is the option to decompose between-study and within-study variability. To avoid aggregation bias, some researchers \citep{hua_one-stage_2017,debray_get_2015,donegan_assessing_2012,hong2015incorporation} suggest decomposing the interactions into two sources: individual-level (i.e., within-study effect) and aggregate-level (i.e., between-study effect) interactions. This model can be written by extending Equation (\ref{glmm}): 
\begin{align*}
&g(E(Y_{is}))= \alpha_s + \delta_sA_{is} + \boldsymbol{\beta}_{s,\text{within}}^T (\boldsymbol{X}_{is} - \bar{\boldsymbol{X}}_s) +\\ & \boldsymbol{\beta}_{\text{across}}^T \bar{\boldsymbol{X}}_s + \boldsymbol{\theta}_{s, \text{within}}^TA_{is}(\boldsymbol{Z}_{is}-\bar{\boldsymbol{Z}}_s) + \boldsymbol{\theta}_{\text{across}}^TA_{is}\bar{\boldsymbol{Z}}_s.
\end{align*}
Here, we have broken up the covariate and treatment-covariate interaction terms into within-study effect and between-study components so that we can separately assess the associations of individual covariates and their study-level summaries with the outcome. This is especially helpful when specific effect moderators vary significantly both within studies and across studies \citep{debray_get_2015}. Equation (\ref{glmm}) is a special case of this model when $\beta_{\text{across}}$ and $\theta_{\text{across}}$ are equal to the average of the $\beta_{s, \text{within}}$'s and the $\theta_{s, \text{within}}$'s, respectively \citep{hua_one-stage_2017}.

Standard implementations of meta-analysis techniques to assess effect heterogeneity assume that a set of potential moderators has already been identified and observed in all included studies. Because studies measure several variables that could plausibly serve as effect moderators, selecting which terms to include in the model is an important and challenging decision. Furthermore, testing a high number of potential effect moderators can increase the risk of false positives \citep{haywardinstrument}. When many potential moderators exist, variable selection or shrinkage methods can help overcome these challenges and identify meaningful moderators while controlling for overfitting. \citeauthor{seo_comparing_2021} (\citeyear{seo_comparing_2021}) compared one-stage IPD meta-analysis methods that identified effect moderators and estimated their effect size. They compared various variable selection methods under both frequentist and Bayesian frameworks including stepwise selection, Lasso regression, Ridge regression, adaptive Lasso, Bayesian Lasso, and stochastic search variable selection (SVSS). In extensive simulation studies, the shrinkage methods (Lasso, Ridge, adaptive Lasso, Bayesian Lasso, and SVSS) performed best, supporting the usage of such methods in IPD meta-analysis to enhance performance \citep{seo_comparing_2021}. Especially in settings in which large numbers of variables are available and many could plausibly serve as treatment effect moderators, these methods could be useful to efficiently estimate the conditional average treatment effect.

\subsubsection{Integrating IPD with AD}

If data are available at the individual level in some studies but at the aggregate level in others, both levels of data can still be combined to estimate treatment effects. One straightforward way to do so is through two-stage meta-analysis, as introduced in \ref{types}, where models are fit to each study with IPD to calculate aggregate statistics, and then these statistics can be combined with those reported in the AD \citep{riley2008meta}. Another more complicated but effective approach is to combine the IPD and AD simultaneously in one-stage meta-analysis: \citeauthor{riley2008meta} (\citeyear{riley2008meta}) describe a method for doing this where the outcome for each trial with only AD is simply the estimate of the treatment effect and there is just one observation. They also incorporate an indicator of IPD versus AD.

Bayesian methodology can also be incorporated to combine IPD with AD and allow for adaptive borrowing of information. In such a setting, \citeauthor{hong2018power} (\citeyear{hong2018power}) recommend treating the AD as auxiliary data and utilizing a power prior to adaptively incorporate the AD and a commensurate prior to borrow from the AD to estimate treatment effects. In another Bayesian approach, \citeauthor{saramago2012mixed} (\citeyear{saramago2012mixed}) incorporate IPD-level covariates to improve estimation of treatment-covariate interactions over that available by AD alone.

\subsection{Combining an RCT with Observational Data} \label{RCTobs}

Another usage for IPD in estimating treatment effect heterogeneity is through combining data from an RCT with an observational dataset. For example, we can consider the scenario introduced earlier where we are interested in comparing two treatments for major depression, duloxetine and vortioxetine, and we have access to RCT data and a large observational dataset containing electronic health records \citep{combining_present}. This scenario requires attention to potential confounding in the observational dataset; notably, the individuals are not randomly assigned to treatment in the observational data unlike in the RCTs. In this setting, the approaches are often nonparametric, with some exceptions, and they include some approach for accounting for confounding in the observational dataset. We use $\hat{\tau}^r(\boldsymbol{X})$ and $\hat{\tau}^o(\boldsymbol{X})$ to represent the estimated CATE function based on data from the RCT and observational study, respectively. 

\cite{colnet_causal_2021} provides a literature review of methods that combine RCT and observational data. They touch on many different purposes of combination, one of which is CATE estimation. Their review includes some of the nonparametric approaches listed in this section \citep{kallus_removing_2018, yang_improved_2022, yang_elastic_2020} and discusses key assumptions, code, and implementation of methods. Our review incorporates some of the same papers but includes other recent and related approaches as well.

Existing methods for combining RCT and observational data first involve estimating the CATE in either the randomized trial data, the observational data, or both, using single-study methods. These estimators are then combined in one of multiple different ways.

\subsubsection{Combining Separate CATE Estimates from RCT and Observational Studies}

When combining one RCT with one large observational dataset (the usual approach in the methods to follow), one category of approaches involves estimating the CATE in both datasets. In several of these approaches, the final CATE estimate is a weighted combination of the two study-specific CATE estimates, where the weight is derived based on a method-specific estimate of bias in the observational data. This is the approach taken by Rosenman et al. in two papers (\citeyear{rosenman2022propensity}; \citeyear{rosenman_combining_2020}). In each paper, Rosenman and colleagues discuss the CATE in terms of average treatment effects within ``strata'', or subgroups that can be defined as a complex function of covariates \citep{rosenman2022propensity}. The authors construct strata based on effect moderators and propensity score estimates from the observational data. They assume that within each stratum, the true average treatment effect is the same for both the observational and RCT data; however, the observational data may yield a biased estimate due to unobserved confounding. The base estimator used in their papers is a difference in mean outcomes between the treatment and control group within stratum $k$: \begin{equation}\label{rosenman}
    \hat{\tau}_{k}^o = \frac{\sum_{i\in O_k}A_iY_i}{\sum_{i\in O_k}A_i} - \frac{\sum_{i\in O_k}(1-A_i)Y_i}{\sum_{i\in O_k}(1-A_i)}
\end{equation} where $o$ indicates observational study, $k$ indexes strata, and $O_k$ is the set of individuals in the observational study belonging to stratum $k$. The same estimator can be established for the RCT by replacing $o$ and $O_k$ with $r$ and $R_k$, respectively. From this, Rosenman et al. (\citeyear{rosenman2022propensity}) construct a ``spiked-in'' estimator, in which individuals from the RCT are assigned to their corresponding strata with individuals from the observational data. Then the stratum-specific treatment effects are estimated as in Equation (\ref{rosenman}) but including both RCT and observational data. They compare this ``spiked-in'' estimator with a dynamic weighted average in which stratum-specific treatment effects are estimated separately in the RCT and observational data, and then the weight for combining the RCT and observational stratum-specific treatment effects is constructed based on the variance of the RCT estimator and the mean squared error (MSE) of the observational data estimator. Ultimately, they discover that the ``spiked-in'' estimator is only effective when the covariate distributions are very similar across datasets and that their dynamic weighted average has low bias regardless of whether the covariate distributions are similar or not.

In their second paper in this stratum-specific treatment effect framework, Rosenman et al. (\citeyear{rosenman_combining_2020}) utilize shrinkage estimation to combine CATE estimators from the RCT and observational dataset. They first determine a structure for a given shrinkage factor, $\lambda$, and then optimize an unbiased risk estimate to solve for this $\lambda$. They again define stratum-specific average treatment effects under the assumption that treatment effect heterogeneity can be assessed by dividing up the dataset into strata. For example, they define a common shrinkage factor $\lambda$ selected by minimizing the unbiased risk estimate such that \begin{equation}\label{rosenman2}
    \hat{\tau}_k(\lambda) = \hat{\tau}_{k}^r - \lambda(\hat{\tau}_{k}^r - \hat{\tau}_{k}^o)
\end{equation} where $r$ indexes the RCT estimator, $o$ the observational estimator, $k$ indexes strata, and $\hat{\tau}_{k}^r$ and $\hat{\tau}_{k}^o$ can be estimated as specified in Equation (\ref{rosenman}). They also discuss an estimator that is the same as Equation (\ref{rosenman2}) but multiplies the difference $\lambda(\hat{\tau}_{k}^r - \hat{\tau}_{k}^o)$ by the variance matrix from the RCT. Note that both of these approaches by Rosenman and colleagues are technically at the subgroup-level; however, these subgroups can be complex functions of covariates, so the approach can be easily discussed in terms of covariates, $\boldsymbol{X}$, instead of stratum membership.

A recent paper by \citeauthor{cheng_adaptive_2021} (\citeyear{cheng_adaptive_2021}) incorporates a similar approach to the shrinkage estimation by Rosenman et al. (\citeyear{rosenman_combining_2020}) by adaptively combining CATE functions between an RCT and observational dataset based on the estimated degree of bias in the observational estimator to yield study-specific CATE estimates that minimize MSE. \citeauthor{cheng_adaptive_2021} (\citeyear{cheng_adaptive_2021}) also use a weighted linear combination of CATE estimators from the RCT, $\hat{\tau}_s^r(\boldsymbol{X})$ and the observational data, $\hat{\tau}_s^o(\boldsymbol{X})$: $$\hat{\tau}_s(\boldsymbol{X}) = \hat{\tau}_s^r(\boldsymbol{X}) + \nu_{\boldsymbol{X}}\{\hat{\tau}_s^o(\boldsymbol{X})-\hat{\tau}_s^r(\boldsymbol{X})\}$$ where $s=0,1$ denotes RCT and observational data, respectively and $\nu_{\boldsymbol{X}}$ is a weight function. To estimate CATE functions in each study separately, the authors use doubly-robust pseudo-outcomes \citep{kennedy_optimal_2020} that are defined as influence functions for the average treatment effect (see more in the Supplementary Material \citep{brantnersupp}). These influence functions are then regressed on the potential effect moderators, $\boldsymbol{X}$, to estimate the CATE in both the RCT ($\hat{\tau}_s^r(\boldsymbol{X})$) and observational data ($\hat{\tau}_s^o(\boldsymbol{X})$) separately. The weight $\nu_{\boldsymbol{X}}$ is estimated by minimizing a decomposition of an estimate of the mean squared error (MSE) for the CATE function and varies based on $\boldsymbol{X}$. This strategy allows for the weight to heavily favor the RCT estimator when the observational data is biased and to combine both estimators efficiently to minimize asymptotic variance in the presence of insignificant bias in the observational data.

Cheng and Cai's method of estimating $\nu_{\boldsymbol{X}}$ is similar to \citeauthor{rosenman_combining_2020} (\citeyear{rosenman_combining_2020}) approach of estimating $\lambda$ using an unbiased risk estimate. An important distinction between the two approaches is that \citeauthor{rosenman_combining_2020} (\citeyear{rosenman_combining_2020}) represent treatment effect heterogeneity through $K$ distinct strata within which they assume that the treatment effect is common across the RCT and observational datasets. Cheng and Cai (\citeyear{cheng_adaptive_2021}) instead use individual covariates as part of their CATE estimation, and they do not require the treatment effects to be equivalent between the RCT and observational datasets. Cheng and Cai (\citeyear{cheng_adaptive_2021}) also use a different base estimation procedure for the initial estimates of $\tau$ in the RCT and observational data.

Finally, \citeauthor{yang_elastic_2020} (\citeyear{yang_elastic_2020}) also combine separate estimates of the CATE from the RCT and observational data to minimize MSE under the assumptions of unconfoundedness in the RCT (Assumption \ref{unconf} in the RCT; satisfied via randomization) and a structural model for the CATE ($\tau(\boldsymbol{X}) = \tau_{\psi_0}(\boldsymbol{X})$). This approach uses elastic integration to combine the estimates based on a hypothesis test that determines whether the assumption of unconfoundedness in the observational data (Assumption \ref{unconf} in the observational data) is sufficiently met or not \citep{yang_elastic_2020}. To construct this test, Yang et al. (\citeyear{yang_elastic_2020}) introduce \begin{equation} \label{yangeq}
    H_{\psi_0}(\boldsymbol{X}) = Y-\tau_{\psi_0}(\boldsymbol{X})A
\end{equation} such that $E(H_{\psi_0}|A,\boldsymbol{X},S) = E(Y(0)|A,\boldsymbol{X},S)$. From here, they introduce a semiparametric efficient score of the parameters $\psi_0$ which we will call $\mathrm{SES}_{\psi_0}$. This semiparametric efficient score is used in their hypothesis test with a null hypothesis of $E(\mathrm{SES}^o_{\psi_0}) = 0$ where $\mathrm{SES}^o_{\psi_0}$ is the score in the observational data. If this null hypothesis is rejected, the ultimate parameters for the CATE are determined solely from the RCT data; if not, parameters are solved for using an elastic integration of both the RCT and observational data. Estimating the parameters is discussed in more detail in Yang et al.'s (\citeyear{yang_elastic_2020}) paper; briefly, they solve $$\frac{\sum_{i=1}^N{\widehat{\mathrm{SES}}}_\psi}{N} = 0$$ by plugging in estimators of unknown quantities and solving for $\psi$.

\subsubsection{Estimating and Accounting for the Confounding Bias in the Observational Data}
Another category focuses on estimating the CATE -- and the confounding bias, as estimated by bringing in the RCT data -- in the observational data, rather than estimating the CATE in each dataset. Kallus and colleagues (\citeyear{kallus_removing_2018}) estimate the CATE in the observational data first and then estimate a correction term to adjust for confounding. They focus on deriving a CATE estimator that is consistent. The approach assumes unconfoundedness (Assumption \ref{unconf}) in the RCT, but does not assume that the observational data fully overlaps with the RCT data \citep{kallus_removing_2018,colnet_causal_2021}. The authors note that the CATE function in the observational data, $\tau^o(\boldsymbol{X})$ does not equal the true CATE, $\tau(\boldsymbol{X})$ because of confounding, so they define the confounding effect to be $$\eta(\boldsymbol{X}) = \tau(\boldsymbol{X}) - \tau^o(\boldsymbol{X})$$ and focus on estimating this $\eta$ to correct the observational CATE estimator. The observational CATE is estimated using any single-study approach, such as a causal forest \citep{athey2019generalized,brantnersupp}, and the confounding effect is estimated using the following equation. For the propensity score in the RCT, $\pi^r(\boldsymbol{X}) = P(A = 1|\boldsymbol{X}, S=0)$, Kallus et al. define $$q^r(\boldsymbol{X}_i) = \frac{A_i}{\pi^r(\boldsymbol{X}_i)} - \frac{1-A_i}{1-\pi^r(\boldsymbol{X}_i)}$$ for individuals in the RCT.
This leads to the final equation to solve to estimate the confounding effect: $$\hat{\boldsymbol{\theta}} = \mathrm{argmin}_{\boldsymbol{\theta}}\sum_{i=1}^{n^r}(q^r(\boldsymbol{X}_i)Y_i-\hat{\tau}^o(\boldsymbol{X}_i)-\boldsymbol{\theta}^T\boldsymbol{X}_i)^2$$ again applied to only individuals in the RCT, where $n^r$ is the total number of individuals in the RCT.
Finally, they set $\hat{\eta}(\boldsymbol{X}) = \hat{\boldsymbol{\theta}}^T\boldsymbol{X}$ and ultimately define $$\hat{\tau}(\boldsymbol{X}) = \hat{\tau}^o(\boldsymbol{X}) + \hat{\eta}(\boldsymbol{X}).$$ 

\citeauthor{yang_improved_2022} (\citeyear{yang_improved_2022}) also estimate confounding in the observational study directly. They focus on the conditional average treatment effect on the treated (CATT), $\tau(\boldsymbol{X}) = E[Y(1)-Y(0)|\boldsymbol{X},A=1]$, and define a confounding function to estimate the effect of unobserved confounding in the observational data. They assume unconfoundedness in the RCT (Assumption \ref{unconf}), a structural model for both the CATT and the confounding function, $\zeta$, and that the RCT and observational data come from the same target population, though their covariate distributions need not overlap. Their confounding function is defined in the observational study as the difference in potential outcome means between treatment groups: \begin{align*}
    \zeta(\boldsymbol{X}) =& E[Y(0)|A=1,\boldsymbol{X},S=1]-\\& E[Y(0)|A=0,\boldsymbol{X},S=1].
\end{align*} When all confounders are measured, $\zeta(\boldsymbol{X})=0$, but in reality, unobserved confounders will lead the function to be non-zero. \citeauthor{yang_improved_2022} (\citeyear{yang_improved_2022}) show that this function is only identifiable when the RCT data is used with the observational data. 

To estimate the parameters for the CATT and the confounding function, \citeauthor{yang_improved_2022} (\citeyear{yang_improved_2022}) utilize estimating equations and semiparametric efficiency theory, similar to the approach taken by \citeauthor{yang_elastic_2020} (\citeyear{yang_elastic_2020}). Specifically, they define an equation similar to that of their previous work \citep{yang_elastic_2020} shown in Equation (\ref{yangeq}): $$H_{\psi_0} = Y-\tau_{\varphi_0}(\boldsymbol{X})A-S\zeta_{\phi_0}(\boldsymbol{X})(A-e(\boldsymbol{X},S))$$ where $\psi_0 = (\varphi_0, \phi_0)$ are parameters and such that the final term in the equation will only come into play when $S=1$, i.e., in the observational data. They solve an estimating equation based around this $H$ to get a preliminary estimator of the parameters for $\tau$ and $\zeta$; next, they update this solution based on a semiparametric efficient score. The authors finally show that their estimator of the CATT, which integrates both datasets, is more efficient than the CATT from the RCT data when the predictors from the CATT function and confounding function are linearly independent.

The ``integrative R-learner'' falls in a similar category of methods and is based on adapting the original R-learner by \citeauthor{nie_quasi-oracle_2021} (\citeyear{nie_quasi-oracle_2021}) (see Supplementary Material \citep{brantnersupp}) to the setting with one RCT and one observational dataset \citep{wu2021integrative}. This approach minimizes loss and is consistent and asymptotically efficient compared to an RCT-only estimator. The authors use a very similar definition of the confounding function as in \cite{yang_improved_2022}, with a slight adjustment: \begin{align*}
    c(\boldsymbol{X}) =& E(Y|\boldsymbol{X},A=1,S=1)-\\&  E(Y|\boldsymbol{X},A=0,S=1)-\tau(\boldsymbol{X})
\end{align*} where $c(X) = 0$ when there is no unobserved confounding in the observational dataset (Assumption \ref{unconf}). \citeauthor{wu2021integrative} (\citeyear{wu2021integrative}) estimate this confounding function and $\tau(\boldsymbol{X})$ by minimizing an empirical loss function that has the Neyman orthogonality property, as found in the original R-learner \citep{nie_quasi-oracle_2021}.

Finally, \citeauthor{hatt2022combining} (\citeyear{hatt2022combining}) propose a method that utilizes the estimated confounding effect in the observational data through a representation learning approach. Under similar assumptions to previous methods such as consistency (Assumption \ref{consistency}), common support across the RCT and observational data (Assumption \ref{studypositivity}), and unconfoundedness in the RCT (Assumption \ref{unconf}) among others, \citeauthor{hatt2022combining} (\citeyear{hatt2022combining}) define $\phi^*$ to be a representation of the shared structure of covariates in both the RCT and the observational data. They also define $h_a^r$ and $h_a^o$ as ``hypotheses'' in the RCT and observational data, respectively, for $a = 0,1$ indicating control or treatment. These so-called hypotheses are functions meant to be applied to the representation, $\phi^*$ where for $r$ representing membership in the RCT and $o$ in the observational data,
\begin{align*}
    &E(Y^r|A^r=a, \boldsymbol{X}^r=\boldsymbol{x}) - E(Y^o|A^o=a, \boldsymbol{X}^o=\boldsymbol{x}) \\ &= h_a^r(\phi^*(\boldsymbol{x})) - h_a^o(\phi^*(\boldsymbol{x})).
\end{align*} Similarly to previous methods, \citeauthor{hatt2022combining} (\citeyear{hatt2022combining}) use a confounding function to represent the bias, defined as $\gamma_a = h_a^r - h_a^o$. Their algorithm starts by estimating $\hat{\phi}$ and $\hat{h}_a^o$ for $a=0,1$ from the observational data by minimizing an empirical loss. Next, these estimates are applied to the RCT data and the empirical loss in this dataset is minimized to derive an estimate for the bias $\hat{\gamma}_a$, $a=0,1$. Finally, these estimates are combined using the fact that $\gamma_a = h_a^r - h_a^o$ to solve for $\hat{h}_a^r = \hat{\gamma}_a + \hat{h}_a^o$ and to ultimately estimate the CATE as $$\hat{\tau}(\boldsymbol{X}) = \hat{h}_1^r(\hat{\phi}(\boldsymbol{X})) - \hat{h}_0^r(\hat{\phi}(\boldsymbol{X})).$$

\section{Discussion} \label{futuredir}

\subsection{Comparison of Approaches}

The recent influx of interest in studying treatment effect heterogeneity has led to novel and adapted methods that strive to improve the identification of tailored interventions. Furthermore, with the increase of IPD availability and the simultaneous research interests of combining data sources, assessing treatment effect heterogeneity in a reproducible manner is more feasible than before. Table~\ref{comparison} summarizes the aforementioned approaches, with a focus on their data setting, modeling approach, and motivation.

\begin{table*}
\caption{Comparison of Approaches to Estimate CATE Using Multiple Studies}
\label{comparison}
\begin{tabular}{
  p{\dimexpr.26\linewidth-3\tabcolsep-1\arrayrulewidth}
  p{\dimexpr.1\linewidth-3\tabcolsep-1\arrayrulewidth}
  p{\dimexpr.15\linewidth-3\tabcolsep-1\arrayrulewidth}
  p{\dimexpr.15\linewidth-3\tabcolsep-1\arrayrulewidth}
  p{\dimexpr.15\linewidth-3\tabcolsep-1\arrayrulewidth}
  p{\dimexpr.3\linewidth-3\tabcolsep-1\arrayrulewidth}
  }
\hline
Approach
& \multicolumn{1}{c}{Data Level}
& \multicolumn{1}{c}{Data Types}
& \multicolumn{1}{c}{Framework}
& \multicolumn{1}{c}{Estimand}
& \multicolumn{1}{c}{Motivation} \\
\hline
Meta-Analysis of Interactions & AD & RCTs  & Parametric & Pooled & Pool treatment-covariate interactions\\
Meta-Regression    & AD & RCTs & Parametric & Pooled &  Model group-level treatment-covariate interactions\\
Meta-Analysis of Local Models & FL & RCTs & Parametric & Pooled & Pool treatment-covariate interactions\\
\cite{tan_tree-based_2021} & FL  & RCTs   & Non-parametric & Study-specific & Borrow information from other studies to improve model\\
One-Stage Meta-Analysis & IPD  & RCTs & Parametric & Pooled & Model individual-level treatment-covariate interactions\\
Meta-Analysis of IPD and AD & IPD/AD & RCTs & Parametric & Pooled & Adaptively incorporate AD as auxiliary data \\
\cite{rosenman2022propensity} & IPD & RCT and OD &  Parametric & Pooled & Weight combination of CATE estimators based on OD bias\\
\cite{rosenman_combining_2020} & IPD & RCT and OD &  Parametric & Pooled & Weight combination of CATE estimators based on OD bias\\
\cite{cheng_adaptive_2021}    & IPD  & RCT and OD  & Non-parametric & Study-specific &   Weight combination of CATE estimators based on OD bias \\
\cite{yang_elastic_2020} & IPD & RCT and OD  & Parametric & Pooled & Weight combination of CATE estimators based on OD bias \\
\cite{kallus_removing_2018}  & IPD & RCT and OD & Non-parametric & Pooled &  Estimate confounding function \\
\cite{yang_improved_2022}   & IPD  &RCT and OD &  Parametric & Pooled & Estimate confounding function\\
\cite{wu2021integrative} & IPD & RCT and OD  & Non-parametric & Pooled & Estimate confounding function\\
\cite{hatt2022combining} & IPD & RCT and OD & Non-parametric & Pooled & Estimate confounding function \\
\hline
\multicolumn{6}{l}{\small *AD: aggregate-level data, FL: federated learning, IPD: individual participant-level data, RCT: randomized controlled trial, OD: observational data}
\end{tabular}
\end{table*}

\subsection{Parametric and Nonparametric Approaches}

Meta-analyses have been in use for many years but are less often conceptualized in terms of identifying treatment effect moderation. This review and some other continuing work (i.e., \citeauthor{seo_comparing_2021}, \citeyear{seo_comparing_2021}) have tied meta-analyses into this framework. Traditional methods for assessing moderation generally have involved parametric approaches that require pre-specification of the potential moderators. However, parametric regression models are limited by the need to pre-specify interaction terms, and complex non-linearities might be missed in the ultimate CATE function. Variable shrinkage techniques (including priors) could help to ensure that the most important interactions are included without overfitting the model \citep{seo_comparing_2021}.

Newer approaches listed in Section \ref{RCTobs} include flexible machine learning methods that allow for complicated functional forms for the covariates in the CATE and do not require that moderators be pre-specified. The nonparametric side to estimation that is often employed when combining an RCT with observational data allows for the CATE function to be more complex, but there are some potential weaknesses of these methods compared with simpler parametric models. First, the resulting CATE estimates may be more difficult to interpret, particularly if the goal is to pick out individual effect moderators and assess their precise relationship with the treatment effect. Second, the desirable theoretical properties of these methods—consistency of the estimators, robustness against model misspecification, accuracy of the associated confidence intervals—are for the most part asymptotic, and so a priori one would expect that the nonparametric/machine learning methods are better suited to situations with enough data. The point at which the robustness of the nonparametric approaches is to be preferred over the explicitness and simplicity of the parametric approaches is perhaps best assessed using a combination of contextual or scientific background knowledge, simulation studies, data splitting techniques like cross-validation and training/test/validation sets, and real-world experience with the methods.

In conclusion, parametric models may suffer from model misspecification but are easy to interpret and apply. Although machine learning methods are relatively untested, their statistical properties are mostly asymptotic, and their implementation can be more computationally intensive, they incorporate a large amount of flexibility and could be ideal when complex nonlinear associations are expected with a large number of variables.

\subsection{Current Shortcomings and Future Directions}

Because this field is growing rapidly and the methods discussed are somewhat new, many methods have not been thoroughly compared to one another in simulation studies or illustrated using real trials and/or observational datasets. There is therefore a broad opening for future research that assesses these approaches in comparison to one another through data applications. For meta-analysis, many real-world applications exist, but not all go in-depth into treatment effect heterogeneity. The remaining approaches discussed in this study are all very recent, and the new methods have not been tried out extensively in real data. Real-world applications will be important for understanding the practical implications and considerations such as differential measurement across datasets, missing data, and more -- such implications must be addressed for the methods to be fully useful in applications. Furthermore, any comparisons that have been done do not combine parametric and nonparametric approaches in this field of CATE estimation using multiple studies.

Another useful field of follow-up study is consolidating and evaluating assumptions. The assumptions of methods discussed here vary in whether they are required, relaxed, or unneeded. It would be helpful to be able to empirically evaluate the assumptions across datasets to examine their feasibility, although not all assumptions explored in this paper can be empirically assessed. Specific approaches for inference in the form of variance estimation and confidence intervals are also needed in many approaches. For parametric approaches discussed throughout the review, often standard methods such as Wald confidence intervals can be employed \citep{yang_improved_2022}, or bootstrapping can be used to estimate intervals and standard errors as well. However, there is an opening for more work to determine the best inference approaches in the parametric and nonparametric cases, and how these approaches vary depending on the method. 

More work could also be done when it comes to the type of data being combined. One might be interested in determining how to apply the meta-analytic framework to the combination of trial and observational data; this field has been called cross-design synthesis and has been debated in the literature \citep{debray_get_2015}. On the other hand, the methods geared towards combining an RCT with observational data could be tailored to combine multiple RCTs, but this option was not discussed in the methods previously described aside from briefly in the federated learning setting \citep{tan_tree-based_2021}

In terms of specific data availability settings, aggregate-level data consistently provides a challenge for estimating individual-level effect moderation, and there are only a couple of limited settings in which this goal can be achieved. Therefore, more IPD data access is the simplest solution to being able to derive an in-depth model to estimate the CATE. For the case when IPD is available but cannot be shared across studies (i.e., federated learning), the approaches discussed in this review could be tailored to deal with this. Very few methods exist in this field within federated learning; only one paper specifically discusses treatment effect heterogeneity when data is distributed privately across studies \citep{tan_tree-based_2021}. Thus, future work could be done to derive approaches to estimate the CATE in federated learning.

Data availability also can vary within a given set of studies, and researchers often run into the issue of systematically missing covariates – i.e., covariates available in some but not all data sources. Covariates also can be sporadically missing, where the covariate is present in all studies but missing for some individuals throughout the studies. Future development of the methods discussed previously should incorporate these considerations, as many of the new approaches leave this for future work. Some papers have looked into these types of missingness in a slightly separate context \citep{colnet:hal-03473691}; for example, \citeauthor{audigier2018multiple} (\citeyear{audigier2018multiple}) investigated the performance of multiple imputation procedures for systematically and sporadically missing data.  \citeauthor{jolani2015imputation} (\citeyear{jolani2015imputation}) also describe a generalized imputation approach for IPD meta-analysis when covariates are systematically missing.

An appropriate follow-up question from this work is when to best implement each method. Because the machine learning methods have not been compared to one another in simulation studies, it is difficult to conclude which of the methods is optimal in which scenario. This review does attempt to clarify which type of data can be handled by each method, and whether the method works with RCT and observational data, or  multiple RCTs. However, further study is needed to determine which approach will yield the most accurate predictions depending on the types of heterogeneity present in the study (i.e., heterogeneity across studies, heterogeneity within studies).

For those working in this field or those who want to learn more, it is important to continue to look out for new research that comes out, since this field is changing and growing rapidly. At the time of this review, many future directions of work are open for pursuit. The new methods mentioned throughout this review increase the feasibility of reproducible conclusions regarding individualized treatment decisions. Because we can employ data from multiple sources, we are developing a deeper understanding and can more effectively estimate individual treatment effects that are reliable and generalizable.



\section*{Acknowledgements}
The authors would like to thank the anonymous referees and the Editor for their constructive comments that improved the
quality of this paper.

\section*{Funding}

Research reported in this publication was partially funded through a Patient-Centered Outcomes Research Institute (PCORI) Award (ME-2020C3-21145; PI: Stuart) and by the National Institute of Mental Health (R01MH126856; PI: Stuart). The statements in this work are solely the responsibility of the authors and do not necessarily represent the views of the Patient-Centered Outcomes Research Institute (PCORI), its Board of Governors or Methodology Committee, or of the National Institute of Mental Health.

\bibliographystyle{unsrtnat}
\bibliography{Review_Arxiv}

\appendix
\section*{Appendix}
\subsection*{Single-Study CATE Estimation Methods} \label{SSML}

In this section, we review several approaches geared towards CATE estimation in a single RCT or observational dataset when we have access to the individual participant-level data (IPD). One option is through a regression, using $$g(E(Y_i)) = \beta_0 + \beta_aA_i + \boldsymbol{\beta}_x^T\boldsymbol{X}_i + \boldsymbol{\beta}_z^TA_i\boldsymbol{Z}_i,$$ where $\boldsymbol{Z}$ represents effect moderators and is a subset of $\boldsymbol{X}$. Traditionally, regressions like this are used to examine pre-determined subgroups; however, this model can also be used for CATE estimation. Specifically, we can define the CATE from this as $$\tau(\boldsymbol{Z}_i) = \beta_a + \boldsymbol{\beta}_z^T\boldsymbol{Z}_i.$$ This approach is built upon in the IPD meta-analysis framework discussed in the main paper in the section entitled, ``One-Stage IPD Meta-Analysis''. In observational data, we can also incorporate propensity score methods before modeling using the above regression to account for confounded treatment assignment. As another option, $\boldsymbol{Z}$ could be a risk score, rather than a set of effect moderators \citep{kent_assessing_2010,kent2020predictive}. This approach is the same but requires some preliminary modeling to derive a risk score.

There are also several machine learning methods for CATE estimation in the single study setting that are still relatively new. These methods can be grouped into the following four classes. The first relies mainly on modeling the conditional mean of the outcome given covariates under each intervention (treatment and control), with the CATE function taken as the difference between the two conditional mean outcome functions. This can be achieved via two different models fit to the two treatment groups separately, or via a single model fit to the full sample; these two strategies have been labeled ``T-learner'' and ``S-learner'' (with T standing for “two” models and S for “single” model), respectively \citep{kunzel2019metalearners}. Based off of these options, the ``X-learner'' has also been developed. In this approach, one first estimates the conditional mean outcomes in each treatment group, $\mu(\boldsymbol{x},0)=E(Y(0)|\boldsymbol{X}=\boldsymbol{x})$ and $\mu(\boldsymbol{x},1)=E(Y(1)|\boldsymbol{X}=\boldsymbol{x})$. Next, individual counterfactual outcomes for each treatment group are imputed by using outcome estimators fit to individuals from the \emph{other} group Here, the subscript $i: A=1$ refers to individual $i$ from the treatment group, and $i: A=0$ refers to individual $i$ from the control group. \begin{align*}
    &\Tilde{D}_{i: A=1} = Y_{i: A=1}-\hat{\mu}(\boldsymbol{X}_{i: A=1},0)\\ &\Tilde{D}_{i: A=0} = \hat{\mu}(\boldsymbol{X}_{i: A=0},1)-Y_{i: A=0}.
\end{align*} Finally, a CATE estimator is calculated in each treatment group using a regression with $\Tilde{D}_{i: A=1}$ and $\Tilde{D}_{i: A=0}$ as outcomes, respectively; the ultimate CATE estimate is then a weighted average (where weights are often estimates of propensity scores) of the CATE functions from each group \citep{kunzel2019metalearners}. These methods can utilize approaches like random forests \citep{breiman2001random,athey2019generalized} or Bayesian additive regression trees (BART) \citep{chipman2010bart} to perform the first step of outcome function estimation. A helpful review of these and related methods are in \citeauthor{caron2020estimating} (\citeyear{caron2020estimating}).

The second class of single-study methods for estimating treatment effect heterogeneity involves transformation of either the outcome or the covariates. An option for transforming the outcome can be written as $$Y_i^*  = Y_i\frac{A_i}{\pi(\boldsymbol{X}_i)} - Y_i\frac{1-A_i}{1-\pi(\boldsymbol{X}_i)}$$ where $\pi(\boldsymbol{X}_i)$ are the propensity scores (probability of treatment assignment given covariates) \citep{signoro2007,powers2018some}. Since $E(Y_i^*|\boldsymbol{X}_i) = \tau(\boldsymbol{X}_i)$, a regression model can be fit to this transformed outcome to estimate $\tau(\boldsymbol{X})$, obviating the need to model outcome mean functions (treated as ``nuisance'' parameters). Transformation of the covariates is another option; this transformation is often via some sign flipping and scaling so that the systematic part of the model fit to the transformed variables represents treatment effect variation, while variation in the mean outcome that is unrelated to treatment effect is relegated to the error part of the model. The modified covariate method (MCM) \citep{tian2014simple} is an example of this approach. 

The third class similarly includes transformation of the covariates but still estimates nuisance parameters in the process. The ``R-learner'' \citep{nie_quasi-oracle_2021} is one example, which first estimates conditional mean outcomes and propensity scores and then uses those estimates in a ``quasi-oracle'' objective function that is optimized. This class also includes transformed outcome methods, where the first step assembles a function, $f(.)$, such that $E(f(.)|\boldsymbol{X}) = \tau(\boldsymbol{X})$, and the second step regresses $f(.)$ on $\boldsymbol{X}$. \citeauthor{kennedy_optimal_2020} (\citeyear{kennedy_optimal_2020}) uses this type of method through an influence function for the average treatment effect. The Bayesian causal forest \citep{hahn2020bayesian} also exists within this class. The Bayesian causal forest parameterizes a function $f$ such that $$f(\boldsymbol{X},\boldsymbol{A}) = \mu(\boldsymbol{X},\boldsymbol{\pi},0) + \tau(\boldsymbol{X})\boldsymbol{A}$$ where $\mu(\boldsymbol{x},\pi,0)=E(Y(0)|\boldsymbol{X}=\boldsymbol{x},\boldsymbol{\pi}=\pi)$, $\boldsymbol{\pi}$ represent the propensity scores, and $\mu(0)$ and $\tau$ have pre-specified prior distributions.

The final class of methods includes trees and forests that partition the covariate space to locally maximize the distance in $\tau(\boldsymbol{X})$ between the sides of each split. Causal inference trees were introduced by \citeauthor{su2012facilitating} (\citeyear{su2012facilitating}) who used recursive partitioning to split data into strata by propensity score and treatment effect. Moving forward from this, causal forests have been developed and then extended to ``honest'' causal forests \citep{athey2019generalized}, wherein for each tree, $Y_i$ can only be used in one of the following: to determine the splitting in the tree, or to estimate the treatment effect within a given leaf \citep{wager2018estimation}. The R-learner approach previously mentioned \citep{nie_quasi-oracle_2021} can also be considered within this class, as it is a causal forest based on residualized exposure and outcome, and the Bayesian causal forest \citep{hahn2020bayesian} is also a part of this class as well.

In all classes, many but not all methods also have built in propensity score-based adjustment for confounding for use in non-randomized studies. A review of several such methods in the observational data setting is by \citeauthor{wendling_comparing_2018} (\citeyear{wendling_comparing_2018}).

\end{document}